\documentclass[12pt]{article}
\usepackage[dvips]{graphicx}
\usepackage{amssymb}
\usepackage{amsmath}
\newcommand{\beq}{\begin{equation}}
\newcommand{\eeq}{\end{equation}}
\newcommand{\bea}{\begin{eqnarray}}
\newcommand{\eea}{\end{eqnarray}}

\def\e2sig{e^{-2r\sigma}}

\begin{document}
\setlength{\baselineskip}{18pt}

\begin{titlepage}

\vspace*{2cm}

\begin{center}
{\Large\bf Flavor Mixing in the Gauge-Higgs Unification\footnote{This is prepared for the proceedings of 
the International Workshop on Grand Unified Theories (GUT2012) held at Yukawa Institute for Theoretical Physics, March 15-17 2012, Kyoto, Japan. 
This presentation was done by N. Maru and 
are based on the works \cite{2010AKLM, 2011AKLM, AKMT}.}} 
\end{center}
\vspace{20mm}

\begin{center}
{\large
Y.~Adachi, N.~Kurahashi$^*$,  C.S.~Lim$^*$, N. Maru$^\dag$ and K.~Tanabe$^*$
}
\end{center}
%
%
%
%
\begin{center}
{\it
Department of Sciences, Matsue College of Technology,
Matsue 690-8518, Japan.}\\
{\it $^*$Department of Physics, Kobe University,
Kobe 657-8501, Japan.} \\
{\it $^\dag$Department of Physics, and Research and Education Center for Natural
Sciences, \\
Keio University, Hiyoshi, Yokohama 223-8521, Japan}\\
\end{center}

%
%
\vspace*{3cm}

\centerline{\large\bf Abstract}
\vspace*{1cm}
Gauge-Higgs unification is the fascinating scenario solving the hierarchy problem without supersymmetry. 
In this scenario, the Standard Model (SM) Higgs doublet is identified with extra component of the gauge field 
in higher dimensions and its mass becomes finite and stable under quantum corrections due to the higher dimensional gauge symmetry. 
On the other hand, Yukawa coupling is provided by the gauge coupling, 
which seems to mean that the flavor mixing and CP violation do not arise at it stands. 
In this talk, we discuss that the flavor mixing is originated from simultaneously non-diagonalizable bulk and brane mass matrices.
Then, this mechanism is applied to various flavor changing neutral current (FCNC) processes 
via Kaluza-Klein (KK) gauge boson exchange at tree level 
and constraints for compactification scale are obtained. 


\end{titlepage}

\section{Introduction}

The hierarchy problem is one of the guiding principle motivating us to go to the physics beyond the SM. 
Among various solutions for it, the gauge-Higgs unification (GHU)  scenario is one of the attractive scenarios 
in which the SM Higgs is identified with extra components of higher dimensional gauge field. 
As the result, the Higgs mass becomes finite and stable under quantum corrections 
due to the higher dimensional gauge symmetry. 

On the other hand, Yukawa coupling is originated from the higher dimensional gauge coupling in the scenario, 
which makes it very hard to realize Yukawa hierarchy, flavor mixing and CP violation 
since the gauge coupling is universal and real.
In this talk, the first two issues are addressed.\footnote{As for the CP violation in GHU, see for instance \cite{CP}.} 
The key ingredients to analyze these issues are the bulk and brane masses. 

\section{Yukawa hierarchy and Flavor mixing}

Let us consider the 5D $SU(3) \times U(1)$ GHU model compactified on an orbifold $S^1/Z_2$ 
whose $Z_2$ parity orbifolding matrix $P={\rm diag}(-,-,+)$.\footnote{The color $SU(3)$ is simply omitted.} 
The gauge symmetry $SU(3)$ is broken to $SU(2) \times U(1)$ by the boundary condition 
and the electroweak symmetry $SU(2) \times U(1)$ is broken to $U(1)_{{\rm em}}$ by Hosotani mechanism. 
The Langrangian for fermion is 
\begin{equation}
{\cal L} = \bar{\Psi}_i iD\!\!\!\!/ \Psi_i - M^{ij} \varepsilon(y) \bar{\Psi}_i \Psi_j
\label{5Dfermion}
\end{equation}
where $\varepsilon(y)$ is a sign function with respect to the coordinate of the fifth dimension $y$, 
$M^{ij}$ is a constant bulk mass matirx and the indices $i,j$ denote generation. 
Note that the bulk mass term is odd under $y \leftrightarrow -y$ from the consistency with $Z_2$ parity symmetry.  

It is easy to obtain zero mode functions for the fermion.
\begin{eqnarray}
f_L^{(0)}(y) = \sqrt{\frac{M^i}{1-e^{-2\pi M^i R}}} e^{-M^i |y|}, \quad
f_R^{(0)}(y) = \sqrt{\frac{M^i}{e^{2\pi M^i R} -1}} e^{M^i|y|}
\end{eqnarray}
where $R$ is the compactification radius and 
the bulk masses are taken to be diagonal in generation space as will be discussed later. 
4D effective Yukawa coupling is given by an overlap integral of zero mode functions 
for fermion\footnote{The zero mode function for Higgs is $y$-independent in flat space case.}
\begin{equation}
Y_{eff} = g_4 \int dy f_L^{(0)}(y) f_R^{(0)}(y) 
\simeq 2\pi M^i R g_4 e^{-2\pi M^i R}
\end{equation}
where $g_4$ is a 4D gauge coupling constant defined by the 5D one $g_5/\sqrt{\pi R}$. 
We immediately see that Yukawa couplings except for top Yukawa coupling can be easily realized 
by ${\cal O}(1)$ tuning of bulk masses $M^i$. 
Top Yukawa coupling can be obtained by embedding top quark into a large dimensional representation with 4 indices 
and taking a vanishing bulk mass \cite{CCC}.

As for the flavor mixing, let us go back to (\ref{5Dfermion}). 
At first sight, the bulk mass matrix can have off-diagonal elements in general, 
therefore it seems to be trivial to realize the flavor mixing. 
However, note that the bulk masses can be always diagonalized by a unitary transformation without changing diagonal kinetic term. 
This means that we have no flavor mixing only in the bulk since Yukawa coupling included in the kinetic term is also diagonal. 
In order to get the quark sector of the SM, we introduce the following 5D fermions \cite{CCC}.
\begin{eqnarray}
\psi^{1,2,3}({\bf 3}) \supset Q^{1,2,3}_{3L}, d_R^{1,2,3}, \quad 
\psi^{1,2}({\overline{{\bf 6}}}) \supset Q_{6L}^{1,2}, u_R^{1,2}, \quad 
\psi^3(\overline{{\bf 15}}) \supset Q_{15L}^3, t_R
\end{eqnarray}
where the numbers in the parenthesis are representations of $SU(3)$ and the SM quarks, 
which are decomposed under $SU(2) \times U(1)$, 
remain as massless zero modes under the $Z_2$ parity. 
Note here that we have two quark doublets per generation. 
Therefore, we identify one of their linear combination with the SM doublet and 
give the other exotic one a mass by the following brane localized mass terms. 
\begin{eqnarray}
{\cal L}_{{brane}} &=& \delta(y) \sqrt{2\pi R} \bar{Q}_R^i(x)
\left[
\eta_{ij} Q_{3L}^j(x,y)
+ \lambda_{ij} Q_L^j(x,y)
\right]
+ \cdots, \\
&=& \delta(y) \bar{Q'}_R^i \left[ m_{{\rm diag}} \quad  {\bf 0}_{3 \times 3} \right]_{ij} 
\left(
\begin{array}{c}
Q_H \\
Q_{SM} \\
\end{array}
\right)_L^j + \cdots
\end{eqnarray}
where 
\begin{equation}
\left(
\begin{array}{c}
Q_3 \\
Q \\
\end{array}
\right)_L
= 
\left(
\begin{array}{cc}
U_1 & U_3 \\
U_2 & U_4 \\
\end{array}
\right)
\left(
\begin{array}{c}
Q_H \\
Q_{SM} \\
\end{array}
\right)_L, \quad U^{\bar{Q}} Q_R =Q_R'. 
\end{equation}
$Q_L^i = (Q^1_{6L}, Q^2_{6L}, Q^3_{15L})^T$, 
$\eta_{ij},\lambda_{ij}$ are $3 \times 3$ matrices. 
$Q_R^i$ are 4D brane localized fields and introduced by hand to give the exotic doublet a mass. 
$\cdots$ means the brane localized mass terms for other exotic fermions \cite{AKMT}. 

Using these relations, 4D Yukawa coupling can be read from the five dimensional gauge coupling as
\begin{eqnarray}
-\frac{1}{2} 
\left(
\langle H^\dag \rangle \bar{d}^{i(0)}_R Y_{eff}^i U_3^{ij} Q_{SML}^{j(0)} 
+ \langle H^t \rangle i \sigma_2 \bar{u}^{i(0)}_R (W Y_{eff})^i U_4^{ij} Q_{SML}^{j(0)}
\right) + h.c.
\end{eqnarray}
where $W \equiv {\rm diag}(\sqrt{2}, \sqrt{2}, 2)$. 
It turns out that the SM Yukawa matrices of up (down) type quark sector are given as
\begin{eqnarray}
\frac{g_4}{2} Y_d = \frac{g_4}{2} Y_{eff} U_3, \quad \frac{g_4}{2} Y_u = \frac{g_4}{2} W Y_{eff} U_4
\end{eqnarray}
and diagonalized by bi-unitary transformations as usual.
\begin{eqnarray}
&&\hat{Y}_d = {\rm diag} 
\left(
\frac{m_d}{m_W}, \frac{m_s}{m_W}, \frac{m_b}{m_W}
\right) = V^\dag_{dR} Y_d V_{dL}, \quad
\hat{Y}_u = {\rm diag}
\left(
\frac{m_u}{m_W}, \frac{m_c}{m_W}, \frac{m_t}{m_W}
\right) = V^\dag_{uR} Y_u V_{uL}, \nonumber \\
&&V_{{}\rm CKM} = V^\dag_{dL} V_{uL}
\end{eqnarray}
We should note that the source of flavor mixing is originated from $3 \times 3$ matrices $U_{3,4}$.

\section{$\Delta F=2$ FCNC}

In this section, we apply the mechanism of flavor mixing discussed in the previous section 
to the various representative $\Delta F=2$ FCNC processes, 
namely $K^0-\bar{K}^0$, $D^0-\bar{D}^0$, $B_d^0-\bar{B}_d^0$ and $B^0_s-\bar{B}^0_s$ mixings \cite{2010AKLM, 2011AKLM, AKMT}. 
In our model, the most dominant contributions to $\Delta F=2$ FCNC are due to KK gluon exchange at the tree level.  

The relevant 4D effective QCD interaction vertices for the down-type quarks are derived as follows. 
\begin{eqnarray}
{\cal L}_s &\supset& \frac{g_2}{2\sqrt{2\pi R}}  
\left(
\bar{d}^i_R G_\mu \gamma^\mu d^i_R + \bar{d}^i_L G_\mu \gamma^\mu d^i_L
\right)
+ \frac{g_s}{2} \bar{d}^i_R G_\mu^{(n)} \gamma^\mu d^j_R 
\left(
V^\dag_{d R} I_{RR}^{(0n0)} V_{d R}
\right)_{ij} \nonumber \\
&&+ \frac{g_s}{2} \bar{d}^i_L G_\mu^{(n)} \gamma^\mu d^j_L (-1)^n 
\left\{
V_{d L}^\dag
\left(
U_3^\dag I_{RR}^{(0n0)} U_3 + U_4^\dag I_{RR}^{(0n0)} U_4
\right)V_{d L}
\right\}_{ij}
\label{QCDint}
\end{eqnarray}
where $I_{RR}^{(0n0)}$ is an overlap integral between zero mode fermions and KK gauge bosons,
\begin{equation}
I_{RR}^{i(0n0)} \equiv \int_{-\pi R}^{\pi R} dy (f_R^i)^2 \cos\frac{n}{R}y 
= \frac{1}{\sqrt{\pi R}} \frac{(2RM^i)^2}{(2RM^i)^2+n^2} \frac{(-1)^ne^{2\pi RM^i}-1}{e^{2\pi RM^i}-1}. 
\end{equation}
As for the up-type quarks, we only have to replace $d \leftrightarrow u$ in (\ref{QCDint}). 
We can see from (\ref{QCDint}) that the flavor mixing appears in the coupling of non-zero KK gluons 
because the overlap integral $I_{RR}^{(0n0)}$ is not proportional to the unit matrix, 
while the coupling of zero mode gluon is flavor conserving due to the flatness of the gluon zero mode function 
and the orthonormality of fermion zero mode functions. 

Using these vertices, we have calculated three types of diagrams (namely LR, LL, RR types depending on the type of currents) 
via KK gluon exchange at the tree level 
corresponding to the dimension six operators with $\Delta F=2$ FCNC. 
%
Comparing the results to the experimental data for the coefficients of the dimension six operators with $\Delta F=2$, 
the lower bounds for the compactification scale have been obtained as follows. 
\begin{eqnarray}
&&R^{-1}  \gtrsim {\cal O}(10)~{\rm TeV}~(K^0-\bar{K}^0),\quad
R^{-1} \gtrsim {\cal O}(1)~{\rm TeV}~(D^0-\bar{D}^0), \\
&&R^{-1} \gtrsim {\cal O}(1)~{\rm TeV}~(B_d^0-\bar{B}_d^0, B_s^0-\bar{B}_s^0). 
\end{eqnarray}
Note that these lower bounds are rather mild compared to what we naively obtain assuming a coefficient of ${\cal O}(1)$ for the dimension six operator,  
namely $R^{-1} \gtrsim {\cal O}(1000)$~TeV. 
This can be understood as follows. 
For the first and second generations, the GIM-like mechanism works since their bulk masses are large to reproduce Yukawa hierarchy 
and the amplitudes behave roughly as ${\rm exp}[-2\pi M^{1,2}R]$. 
For the third generation, while the GIM-like mechanism does not work due to the vanishing bulk mass, 
the amplitudes are suppressed because of the small mixing angles for the 1-3 and the 2-3 generations. 

\section{Summary}
In the gauge-Higgs unification scenario, we have clarified that the flavor mixing is generated 
by the fact that the bulk mass matrix and the brane one cannot be simultaneously diagonalized. 
As an application, we have calculated the representative $\Delta F=2$ processes via KK gluon exchange at the tree level 
and obtained the lower bounds of the compactification scale for corresponding processes. 
The obtained results are rather mild compared to the naive estimation 
because of the GIM-like mechanism for the first, second generations 
and the small mixing of 1-3, 2-3 generations for the third generation.

 \subsection*{Acknowledgments}

We would like to thank the organizers for the opportunity to present our work 
at this stimulating workshop. 
This work was supported in part by the Grant-in-Aid for Scientific Research
of the Ministry of Education, Science and Culture, No.~21244036 and 
in part by Keio Gijuku Academic Development Funds (N.M.).


\end{document}